\documentclass[nobibnotes,preprintnumbers,aps,prl,superscriptaddress,showpacs,twocolumn,twoside,sort&compress]{revtex4}				 %
\usepackage{graphicx}																												 %
\usepackage{titlesec}																											     %
\usepackage{fancyheadings}																							 				 %
\begin{document}																													 %
%%%%%%%%%%%%%%%%%%%%%%%%%%%%																										 %
%																																	 %	
\preprint{{Vol.XXX (201X) ~~~~~~~~~~~~~~~~~~~~~~~~~~~~~~~~~~~~~~~~~~~~~~~~~~~~ {\it CSMAG`16}										  ~~~~~~~~~~~~~~~~~~~~~~~~~~~~~~~~~~~~~~~~~~~~~~~~~~~~~~~~~~~~ No.X~~~~}}																 %			
\vspace*{-0.3cm}																													 %
\preprint{\rule{\textwidth}{0.5pt}}																											 \vspace*{0.3cm}																														 %
%%%%%%%%%%%%%%%%%%%%%%%%%%%%%%%%%%%%%%%%%%%%%%%%%%%%%%%%%%%%%%%%%%%%%%%%%%%%%%%%%%%%%%%%%%%%%%%%%%%%%%%%%%%%%%%%%%%%%%%%%%%%%%%%%%%%%%

\title{Breakdown of a Magnetization Plateau in Ferrimagnetic Mixed Spin-(1/2,$S$) Heisenberg Chains Due to a Quantum Phase Transition Towards the Luttinger Spin Liquid}

% author.  The \affiliation command should follow the
\author{J. Stre\v{c}ka}
\thanks{corresponding author; e-mail: jozef.strecka@upjs.sk}
\affiliation{Institute of Physics, Faculty of Science, P. J. \v{S}af\'arik University, Park Angelinum 9, 040 01 Ko\v{s}ice, Slovakia}

\begin{abstract}
Magnetization curves of the ferrimagnetic mixed spin-(1/2,$S$) Heisenberg chains are calculated with the help of density-matrix renormalization group method for several quantum spin numbers $S$=1, 3/2, 2 and 5/2. It is shown that the ferrimagnetic mixed spin-(1/2,$S$) Heisenberg chains exhibit irrespective of the spin value $S$ exactly one intermediate magnetization plateau, which can be identified with the gapped Lieb-Mattis ferrimagnetic ground state. The magnetization plateau due to the Lieb-Mattis ferrimagnetism breaks down at a quantum phase transition towards the Luttinger spin liquid, which is characterized by a continuous change of the magnetization with the magnetic field until another quantum critical point is reached at the saturation field. 
\end{abstract}

\maketitle

\section{Introduction}

Over the last few years, the ferrimagnetic mixed spin-$s$ and spin-$S$ Heisenberg chains with regularly alternating spins $s=1/2$ and $S>1/2$ have attracted a great deal of attention, since they exhibit a quantum phase transition between intriguing ground states that are manifested in respective magnetization curves as quantized magnetization plateaux and Luttinger spin liquids \cite{yam99,sak99,hon00,yam00,sak02,ten11}. The intermediate magnetization plateaux of the mixed spin-(1/2,$S$) Heisenberg chains should obey the quantization condition known as Oshikawa-Yamanaka-Affleck (OYA) rule $m_s-m$ = integer, where $m_s = S + 1/2$ and $m$ are the total spin and total magnetization per elementary unit \cite{oya97}. According to OYA rule, one of possible ways to increase the total number of magnetization plateaux may consist in increasing size of the constitutent spin $S$. It should be stressed, however, that OYA criterion provides just necessary but not sufficient condition for a presence of a magnetization plateau, whose actual existence has still to be verified by explicit calculations. 

Any bipartite quantum ferrimagnet (irrespective of spin magnitude and spatial dimensionality) should also satisfy the Lieb-Mattis (LM) theorem \cite{lie62}, which assures the following total magnetization $m = S - 1/2$ per unit cell within the zero-field ground state of the ferrimagnetic mixed spin-(1/2,$S$) Heisenberg chains. Hence, OYA criterion in combination with LM theorem would suggest that the ferrimagnetic mixed spin-(1/2,$S$) Heisenberg chains may display one and just one quantized magnetization plateau (regardless of the spin size $S$) at the following fractional value of the total magnetization $m/m_s = (2S-1)/(2S+1)$ normalized with respect to its saturation value. In the present work we will provide a survey for zero-temperature magnetization curves of the ferrimagnetic mixed spin-(1/2,$S$) Heisenberg chains by considering a few different quantum spin numbers $S=1$, $3/2$, $2$ and $5/2$, which will prove all aforementioned features on this paradigmatic class of quantum spin chains.   

\section{Model and method}

Let us consider the mixed spin-$s$ and spin-$S$ quantum Heisenberg chain with regularly alternating spins $s=1/2$ and $S>1/2$ given by the Hamiltonian
\begin{eqnarray}
\hat{\cal H} = J \sum_{j=1}^L \hat{\bf S}_j \cdot (\hat{\bf s}_j + \hat{\bf s}_{j+1}) - h \sum_{j=1}^L (\hat{S}_j^z + \hat{s}_j^z),
\label{ham}
\end{eqnarray}
where $\hat{\bf s}_j \equiv (\hat{s}_j^x,\hat{s}_j^y,\hat{s}_j^z)$ and $\hat{\bf S}_j \equiv (\hat{S}_j^x,\hat{S}_j^y,\hat{S}_j^z)$ denote the usual spin-1/2 and spin-$S$ operators, respectively. The first term entering in the Hamiltonian (\ref{ham}) takes into account the antiferromagnetic Heisenberg interaction $J>0$ between the nearest-neighbor spins and the second term $h = g \mu_{\rm B} H$ incorporating the equal Land\'e g-factors $g_s = g_S = g$ and Bohr magneton $\mu_{\rm B}$ accounts for the Zeemann's energy of individual magnetic moments in an external magnetic field. It is noteworthy that the overall chain length is $2L$ as the elementary unit contains two spins, whereas the translational invariance is ensured by the periodic boundary condition $s_{L+1} \equiv s_1$.

One should turn to some accurate numerical method in order to get a reliable survey of magnetization processes of the ferrimagnetic mixed spin-(1/2,$S$) Heisenberg chains, since the Hamiltonian (\ref{ham}) is not integrable. For this purpose, we have implemented density-matrix renormalization group (DMRG) calculations from ALPS project \cite{bau11}, which can be straightforwardly used to obtain the lowest-energy eigenvalue $E(T_{tot}^z, L, h=0)$ of the ferrimagnetic mixed-spin Heisenberg chain within each sector with the total spin $T_{tot}^z = \sum_{j=1}^L (S_j^z + s_j^z)$ in a zero magnetic field ($h=0$). The lowest-energy eigenstate of the ferrimagnetic mixed spin-(1/2,$S$) Heisenberg chains in a non-zero magnetic field can be subsequently calculated from the formula $E(T_{tot}^z, L, h) = E(T_{tot}^z, L, h=0) - h T_{tot}^z$, because the total spin $T_{tot}^z$ is conserved quantity due to a validity of the commutation relation between the respective operator 
and the Hamiltonian (\ref{ham}). The finite-size formula for a magnetic-field induced transition between the lowest-energy eigenstates with the total spin $T_{tot}^z$ and $T_{tot}^z+1$ then readily follows from the formula $h = E(T_{tot}^z+1, L, h=0) - E(T_{tot}^z, L, h=0)$. In this way one may obtain the accurate numerical results for the zero-temperature magnetization curves. To avoid extrapolation due to finite-size effects we have performed DMRG simulations for a sufficiently large system size with up to $L=64$ units (128 spins), whereas adequate numerical accuracy was achieved through 16 sweeps at the targeted system size when increasing the number of kept states up to 1200 during the final sweeps.

\section{Results and discussion}

\begin{figure}[t]
\includegraphics[width=1.05\columnwidth]{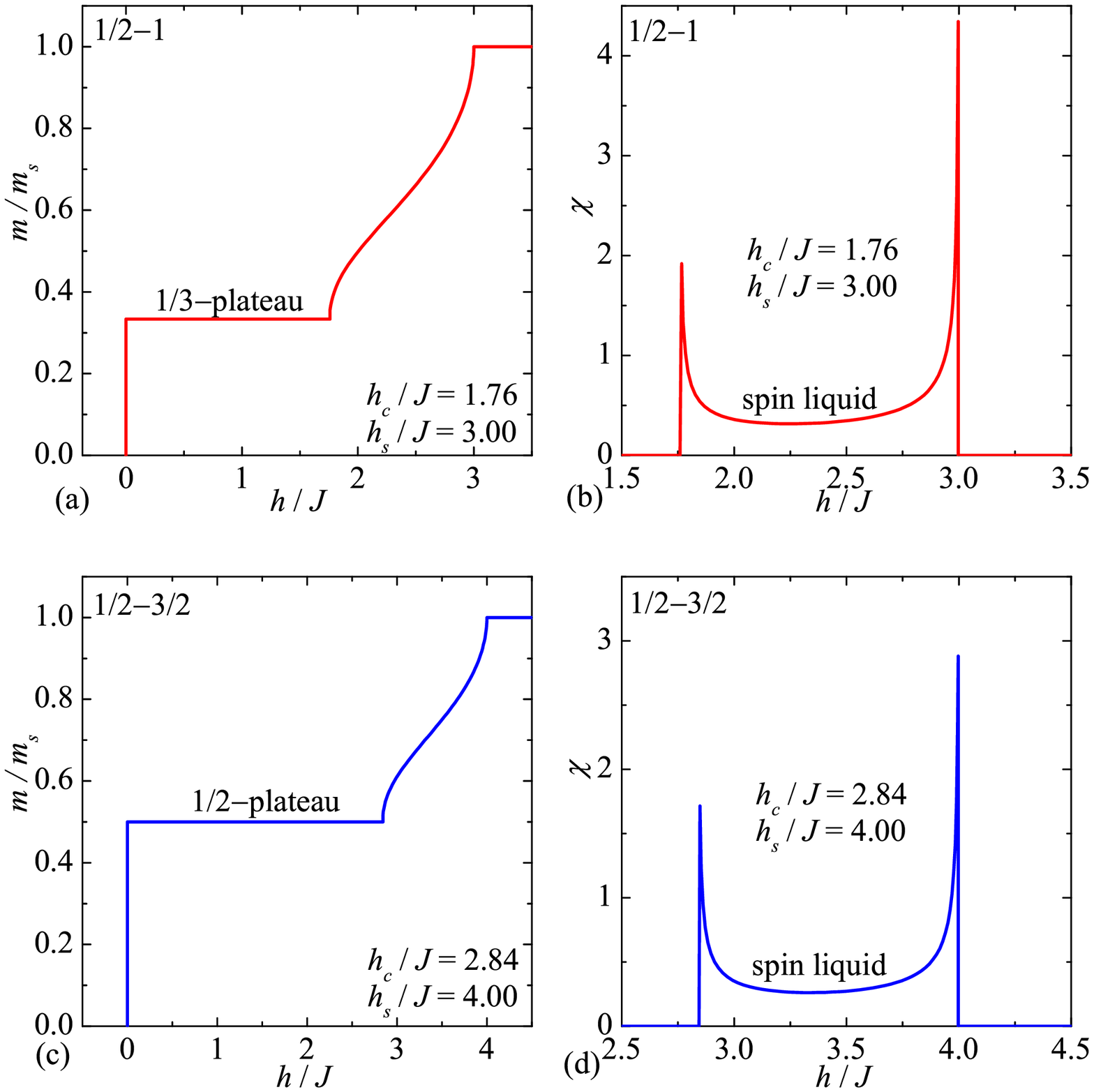}
\includegraphics[width=1.05\columnwidth]{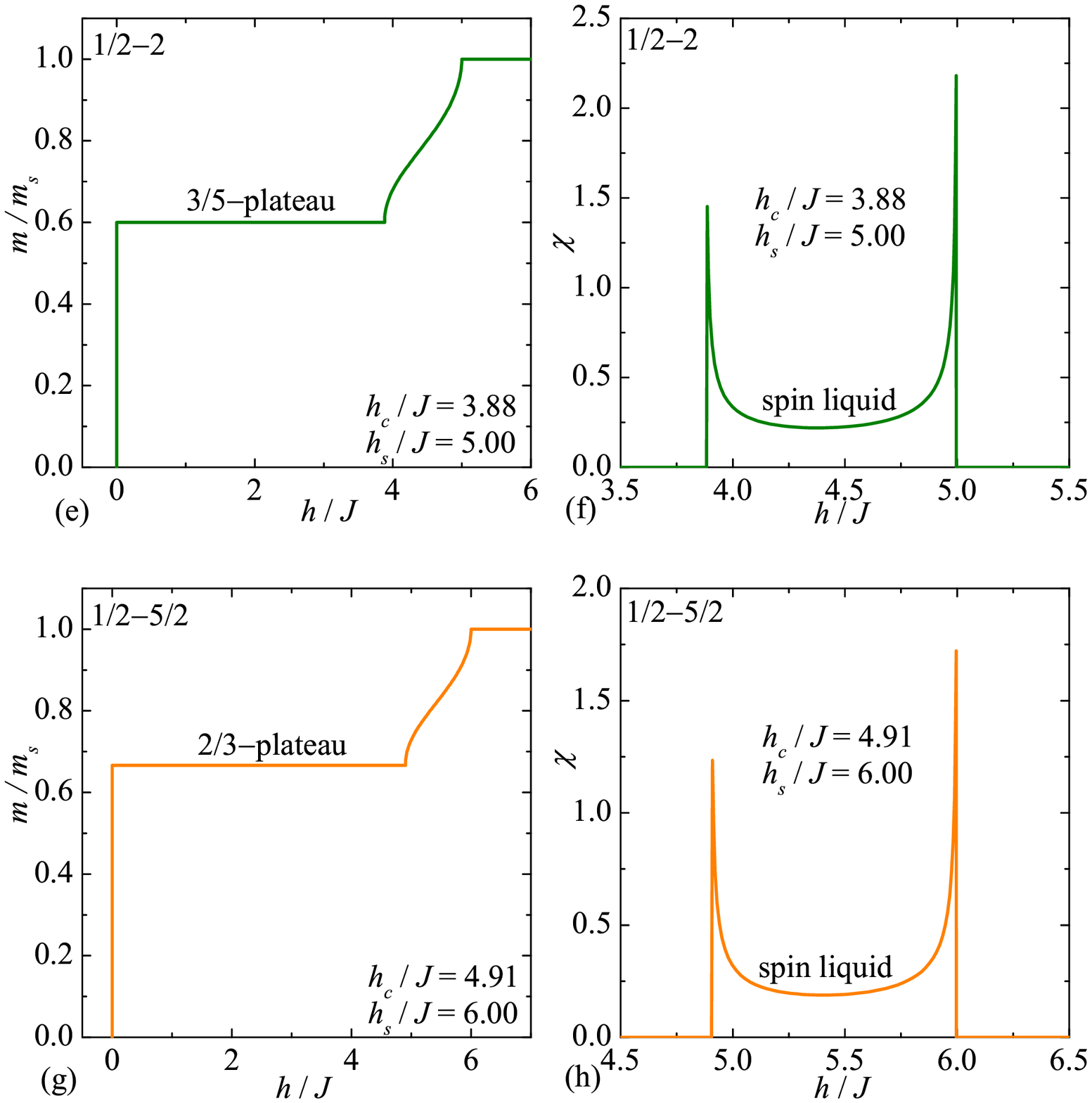}
\vspace{-0.8cm}
\caption{The magnetization (left panel) and susceptibility (right panel) of the mixed spin-(1/2,$S$) Heisenberg chains as a function of the magnetic field for four different spin values: (a)-(b) $S=1$; (c)-(d) $S=3/2$; (e)-(f) $S=2$; (g)-(h) $S=5/2$. The displayed results were obtained from DMRG simulations of a finite-size chain with $L=64$ units (128 spins).}
\label{fig1}
\end{figure}

Let us proceed to a discussion of zero-temperature magnetization curves of the ferrimagnetic mixed spin-(1/2,$S$) Heisenberg chains, which are displayed on the left panel of Fig.~\ref{fig1} for a few different quantum spin numbers $S$=1, 3/2, 2 and 5/2. It is quite evident from Fig.~\ref{fig1} that all considered mixed-spin Heisenberg chains indeed exhibit exactly one intermediate magnetization plateau at the fractional value $m/m_s =(2S-1)/(2S+1)$, which is consistent with the gapped LM ferrimagnetic ground state. The intermediate plateau due to LM ferrimagnetism breaks down at a quantum phase transition invoked by the critical magnetic field $h_c$, which closes an energy gap above the ferrimagnetic ground state. It is noteworthy that the height of LM plateau monotonically increases with increasing the quantum spin number $S$ quite similarly as does its width terminating at the critical field $h_c = 1.76J$ for $S=1$, $h_c = 2.84J$ for $S=3/2$, $h_c = 3.88J$ for $S=2$ and $h_c = 4.91J$ for $S=5/2$. Above the critical magnetic field $h>h_c$ the ferrimagnetic mixed spin-(1/2,$S$) Heisenberg chains pass towards the Luttinger spin liquid, where the magnetization rises continuously with the magnetic field until another quantum critical point is reached at the saturation field $h_s = J(1 + 2S)$. The asymptotic behavior of the magnetization in a vicinity of both quantum phase transitions is governed by the relations: $m \propto \sqrt{h - h_c}$ for $h \to h_{c}^{+}$ and $m \propto \sqrt{h_s - h}$ for $h \to h_{s}^{-}$. Owing to this fact, the quantum phase transitions driven by the magnetic field should be also reflected in anomalous behavior of the magnetic susceptibility $\chi$ close to quantum critical points: $\chi \propto 1/\sqrt{h - h_c}$ for $h \to h_{c}^{+}$ and $\chi \propto 1/\sqrt{h_s - h}$ for $h \to h_{s}^{-}$. In accordance with this statement, the magnetic-field dependences of the susceptibility shown on the right panel of Fig.~\ref{fig1} furnish evidence for both field-induced quantum phase transitions towards the Luttinger spin liquid through the observed divergence of the magnetic susceptibility.

\section{Conclusions}

The zero-temperature magnetization curves of the ferrimagnetic mixed spin-(1/2,$S$) Heisenberg chains were calculated with the help of DMRG method for several values of the quantum spin number $S$. It has been verified that the magnetization curves involve due to the gapped LM ferrimagnetic ground state one and just one intermediate plateau at the fractional magnetization $m/m_s =(2S-1)/(2S+1)$, 
which breaks down at a quantum phase transition towards the Luttinger spin liquid driven by the external magnetic field. Subsequently, the magnetization continuously rises with increasing the magnetic field within the Luttinger spin-liquid phase until it reaches the full moment at the saturation field $h_s = J(1 + 2S)$ closely connected with another field-induced quantum phase transition. It has been demonstrated that the magnetization shows a cusp and susceptibility diverges in a close vicinity of both quantum critical points. Besides, it could be concluded that the rising quantum spin number $S$ increases in the magnetization curve of the mixed spin-(1/2,$S$) Heisenberg chains the height as well as width of the ferrimagnetic LM plateau, while the magnetic-field range corresponding to the gapless Luttinger spin-liquid phase is conversely reduced. Last but not least, it is worth noticing that theoretical implications of the present work are of obvious relevance for series of bimetallic coordination compounds MM'(pba)(H$_2$O)$_3$ $\cdot$ 2H$_2$O \cite{kah87} and  MM'(EDTA) $\cdot$ 6H$_2$O \cite{dri85} (M,M' = Cu, Ni, Co, Mn), which represent experimental realization of the ferrimagnetic mixed-spin Heisenberg chains. However, the high-field magnetization measurements on these or related series of bimetallic complexes are desirable for experimental testing of the present theoretical predictions.        

\section{Acknowledgement}
This work was financially supported by Ministry of Education, Science, Research and Sport of the Slovak Republic provided under the grant No. VEGA 1/0043/16 and by the grant Slovak Research and Development Agency under the contract No. APVV-0097-12.


\begin{thebibliography}{99}
\bibitem{yam99} S. Yamamoto, T. Sakai, \textit{J. Phys.: Condens. Matter} \textbf{11}, 5175 (1999). DOI: 10.1088/0953-8984/11/26/318.
\bibitem{sak99} T. Sakai, S. Yamamoto, \textit{Phys. Rev. B} \textbf{60}, 4053 (1999). DOI: 10.1103/PhysRevB.60.4053.
\bibitem{hon00} A. Honecker, F. Mila, M. Troyer, \textit{Eur. Phys. J. B} \textbf{15}, 227 (2000). DOI: 10.1007/s100510051120.
\bibitem{yam00} S. Yamamoto, T. Sakai, \textit{Phys. Rev. B} \textbf{62}, 3795 (2000). DOI: 10.1103/PhysRevB.62.3795.
\bibitem{sak02} T. Sakai, S. Yamamoto, \textit{Phys. Rev. B} \textbf{65}, 214403 (2002). DOI: 10.1103/PhysRevB.65.214403.
\bibitem{ten11} A.S.F. Ten\'orio, R.R. Montenegro-Filho, M.D. Coutinho-Filho, \textit{J. Phys.: Condens. Matter} \textbf{23}, 506003 (2011). DOI:10.1088/0953-8984/23/50/506003 
\bibitem{oya97} M. Oshikawa, M. Yamanaka, I. Affleck, \textit{Phys. Rev. Lett.} \textbf{78}, 1984 (1997).  DOI: 10.1103/PhysRevLett.78.1984.
\bibitem{lie62} E. Lieb, D. Mattis, \textit{J. Math. Phys.} \textbf{3}, 749 (1962). DOI: 10.1063/1.1724276
\bibitem{bau11} B. Bauer, L.D. Carr, H.G. Evertz, A. Feiguin, J. Freire, S. Fuchs, L. Gamper, J. Gukelberger, E. Gull, S. Guertler, A. Hehn, 
R. Igarashi, S.V. Isakov, D. Koop, P.N. Ma, P. Mates, H. Matsuo, O. Parcollet, G. Pawlowski, J.D. Picon, L. Pollet, E. Santos, V.W. Scarola, 
U. Schollw\"ock, C. Silva, B. Surer, S. Todo, S. Trebst, M. Troyer, M.L. Wall, P. Werner, S. Wessel, \textit{J. Stat. Mech.: Theor. Exp.} \textbf{2011}, P05001 (2011).  
DOI: 10.1088/1742-5468/2011/05/P05001.
\bibitem{kah87} O. Kahn, \textit{Struct. Bonding (Berlin)} \textbf{68}, 89 (1987). DOI: 10.1007/3-540-18058-33. 
\bibitem{dri85} M. Drillon, E. Coronado, D. Beltran, R. Georges, \textit{J. Appl. Phys.} \textbf{57}, 3353 (1985). DOI: 10.1063/1.335094. 
%%%%%%%%%%%%%%%%%%%%%
\end{thebibliography}
\end{document}